\documentclass[twocolumn,showpacs,aps,prl,superscriptaddress]{revtex4}

\usepackage{graphicx}
\usepackage{dcolumn}
\usepackage{amsmath}
\usepackage{epsfig}

\input pubboard/babarsym

\def\mell       {\ensuremath{\ell}\xspace}

\newcommand{\BABARPubYear}    {04}
\newcommand{\BABARPubNumber}  {044}
\newcommand{\SLACPubNumber} {xx}

\def\figurebox#1#2#3{%
    \def\arg{#3}%
    \ifx\arg\empty
    {\hfill\vbox{\hsize#2\hrule\hbox to #2{\vrule\hfill\vbox to #1{\hsize#2\vfill}\vrule}\hrule}\hfill}%
    \else
    {\hfill\epsfbox{#3}\hfill}%
    \fi}

\long\def\inst#1{\par\nobreak\kern 4pt\nobreak
    {\it #1}\par\vskip 10pt plus 3pt minus 3pt}

\begin{document}

\preprint{\babar-PUB-\BABARPubYear/\BABARPubNumber} 
\preprint{SLAC-PUB-\SLACPubNumber} 

\begin{flushleft}
\babar-PUB-\BABARPubYear/\BABARPubNumber\\
% SLAC-PUB-\SLACPubNumber\\
% hep-ex/\LANLNumber\\
\end{flushleft}

\title{
{\large \bf
Measurement of Branching Fractions and Charge Asymmetries for Exclusive \boldmath \B \unboldmath 
Decays to Charmonium
}
}

%% author list as of 02-Nov-2004 (611 authors)
%
\author{B.~Aubert}
\author{R.~Barate}
\author{D.~Boutigny}
\author{F.~Couderc}
\author{Y.~Karyotakis}
\author{J.~P.~Lees}
\author{V.~Poireau}
\author{V.~Tisserand}
\author{A.~Zghiche}
\affiliation{Laboratoire de Physique des Particules, F-74941 Annecy-le-Vieux, France }
\author{E.~Grauges-Pous}
\affiliation{Universitad Autonoma de Barcelona, E-08193 Bellaterra, Barcelona, Spain }
% ...
\collaboration{The \babar\ Collaboration}
\noaffiliation

\date{\today}

\begin{abstract}
We report measurements 
of branching fractions and charge asymmetries of exclusive decays of neutral and 
charged \B mesons into two-body final states containing a charmonium state and a light strange 
meson. The charmonium mesons considered are \jpsi, \psitwos and \chicone, and 
the light meson is either \kaon or \Kstar. We use a sample of about 124 million \BB pairs 
collected with the \babar\ detector at the PEP-II storage ring at the Stanford Linear 
Accelerator Center.
\end{abstract}


\maketitle

Nonleptonic decays of $B$ mesons provide tests of both strong- and weak-interaction dynamics. 
Decays $B\to (c\cbar) K^{(*)}$ are particularly illuminating as they involve three kinds of 
mesons: one with a heavy quark and a light quark, one with two heavy quarks, and one with
two light quarks.  Phenomenological models (see Refs.[2-12] of Ref. \cite{ref:oldanalysis}) 
give estimates for branching fractions and for ratios of the decays to $K$ and $K^*$. 
Some branching fractions and ratios have been reported but others have not, so more stringent tests of 
the models are possible.  The Standard Model predicts small differences between the branching 
fractions for positive and negative $B$ mesons, i.e. small direct CP violation 
\cite{ref:smallasymm}. Large charge asymmetries would be evidence for new physics.    
Limits on direct CP violation would constrain extensions of the Standard Model. Very 
few measurements have been reported in $\B \ra (c \bar{c}) K^{(*)}$ modes \cite{ref:belleca}. 
The decay processes studied in this paper are listed in Table \ref{tab:list}.

\begin{table}[h]
\caption{
Decay processes studied in this analysis. The $(c \bar{c})$ 
resonance is either a \jpsi, \psitwos, or \chicone. For all processes, 
when relevant, we use the secondary decay modes \jpsi,\psitwos \ra \ellell, 
\chicone \ra \jpsi \g, and \KS \ra \pipi, \piz \ra \gaga.}
\begin{center}
\begin{tabular}{l|ll} \hline \hline
  Decay Channel & Secondary \Kstar decay mode                     \\ \hline
  \Bz \ra $(c \bar{c})$ \Kstarz & \Kstarz \ra \Kp \pim , \KS \piz \\
  \Bp \ra $(c \bar{c})$ \Kstarp & \Kstarp \ra \Kp \piz , \KS \pip \\
  \Bz \ra $(c \bar{c})$ \KS     &                                 \\
  \Bp \ra $(c \bar{c})$ \Kp     &                                 \\ \hline
\end{tabular}
\end{center}
\label{tab:list}
\end{table}

The data sample used in this analysis contains 124 million \BB events collected with 
the \babar\ detector at the \pep2\  asymmetric \epem storage ring. This represents a total 
integrated luminosity of 112.4 fb$^{-1}$ taken on the $\Upsilon$(4S) resonance. 
The \babar\ detector is described in detail elsewhere~\cite{ref:babar}. Surrounding the 
interaction point, a five-layer double-sided silicon vertex tracker (SVT) provides 
precise reconstruction of track angles and \B decay vertices. A 40-layer drift 
chamber (DCH) provides measurements of the transverse momenta of charged particles. 
An internally reflecting ring-imaging Cherenkov detector (DIRC) is used for 
particle identification. 
A CsI(Tl) crystal electromagnetic calorimeter (EMC) is used to detect photons and electrons. 
The calorimeter is surrounded by a 1.5-T magnetic field. The flux return is 
instrumented with resistive plate chambers (RPC) used for muon and neutral-hadron identification.

Multihadron events are selected by demanding a minimum of three
reconstructed charged tracks in the polar-angle range $0.41 <\theta < 2.54$\rad, 
where $\theta$ is defined in the laboratory frame. Charged tracks must be reconstructed in the
DCH and are required to originate within 
1.5\cm of the beam in the plane transverse to it and within 10\cm of the beamspot along the 
beam direction. 
Events are required to have a primary vertex 
within 0.5\cm of the average position of the interaction point in the plane transverse to
the beamline, and within 6\cm longitudinally. 
Charged tracks are required to include at least 12 DCH hits and to 
have a transverse momentum \pt$>$100 \mevc. 
Photons are reconstructed from EMC clusters. The lateral energy profile (LAT) \cite{ref:lat} 
is used to discriminate electromagnetic from hadronic clusters. Photons are 
required to have a minimum energy of 30 MeV, to satisfy LAT $<$ 0.8, 
and to be in the fiducial volume $0.41 < \theta < 2.41$ rad. 
Electron candidates are selected using information from the EMC (LAT 
and Zernike moment $A_{42}$ \cite{ref:zernike}), the ratio of the energy measured in the 
EMC to the 
momentum measured by the tracking system, the energy loss in the drift chamber, 
and the Cherenkov angle measured in the DIRC. Electrons are also required to be 
in the fiducial volume $0.41 < \theta < 2.41$ rad. 
Muon candidates are selected using information from the EMC (energy deposition 
consistent with a minimum ionizing particle) and the distribution of hits in the 
RPC. Muons are required to be in the fiducial volume $0.3 < \theta < 2.7$ rad. 
We select charged kaon and pion candidates using information from the 
energy loss in the SVT and DCH, and the Cherenkov angle measured in the DIRC. Kaon 
candidates are required to be in the fiducial volume $0.45 < \theta < 2.45$ rad.

The selection has been 
optimized by maximizing the ratio $S/ \sqrt{S+B}$, where $S$ and $B$ are  
the number of expected signal and background events obtained from Monte Carlo 
simulation. The \jpsi candidates are required to have an invariant 
mass $2.95 < M_{\epem} < 3.14$ \gevcc or $3.06 < M_{\mumu} < 3.14$ \gevcc for \jpsi \ra \epem 
or \jpsi \ra \mumu decays respectively. The \psitwos candidates are required to have an invariant mass 
$3.44 < M_{\epem} < 3.74$ \gevcc or $3.64 < M_{\mumu} < 3.74$ \gevcc for \psitwos \ra \epem 
or \psitwos \ra \mumu decays respectively. For \jpsi,\psitwos\to\epem decays, 
electron candidates are combined with photon candidates in order to recover some of 
the energy lost through bremsstrahlung. 
In the \chicone reconstruction, \jpsi candidates are selected as described above. 
The associated \g has to satisfy LAT $<$ 0.8, $A_{42}<$0.15 
and has to have an energy greater than 0.15 \gev. 
The \chicone candidates are required to satisfy 
$0.35 < M_{\ellell \g} - M_{\ellell} < 0.45$ \gevcc, where \mell represents an electron or a muon. 
The \piz \ra \gaga candidates are required to satisfy $0.113 < M_{\gaga} < 0.153$ \gevcc. 
Both photons have to satisfy LAT $<$ 0.8. The energy 
of the soft photon has to be greater than 0.050 \gev and the energy of the hard 
photon has to be greater than 0.150 \gev. 
The \KS \ra \pip \pim  candidates are required to satisfy $0.489 < M_{\pipi} < 0.507$ \gevcc. 
In addition, the \KS flight distance defined as the distance between the reconstructed \B 
and \KS vertices must exceed 1 mm, and the angle between the \KS momentum and its flight 
direction in the plane transverse to the beam axis must be less than 0.2 rad. 
The \Kstarz and \Kstarp candidates are required to satisfy 
$0.796 < M_{\kaon \pi} < 0.996$ \gevcc and $0.792 < M_{\kaon \pi} < 0.992$ \gevcc, respectively. 
In addition, for the sake of suppressing background from events with soft pions, 
for channels having a \piz in the final state, the cosine of the angle between 
the \kaon momentum and the \B momentum in the \Kstar rest frame has to be less than 0.8.

\begin{figure}
  \center{
  \includegraphics[width=0.15\textwidth]{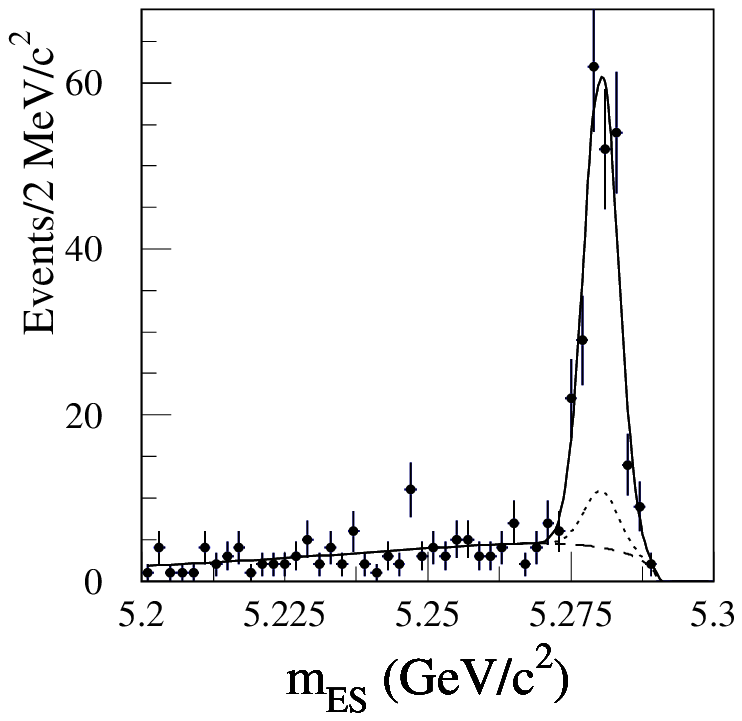}
  \includegraphics[width=0.15\textwidth]{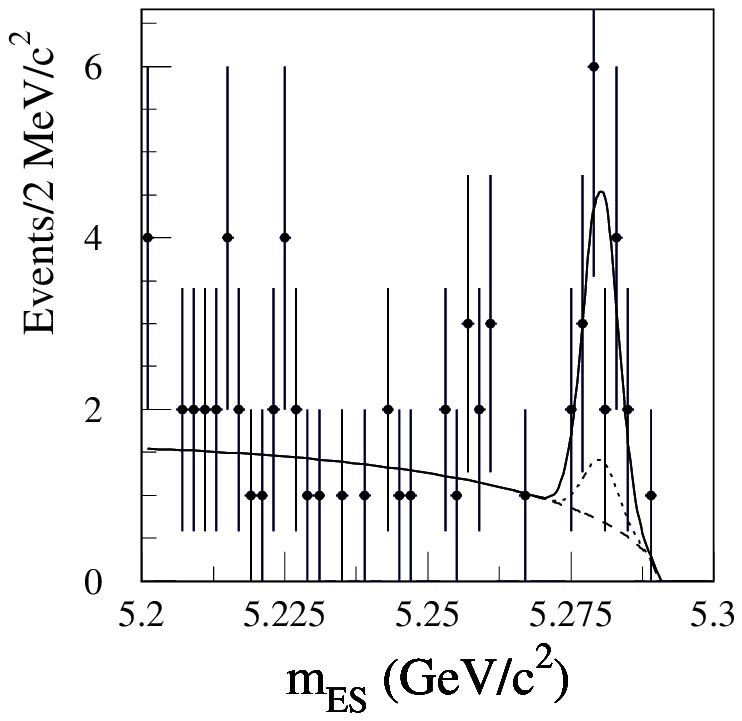}
  \includegraphics[width=0.15\textwidth]{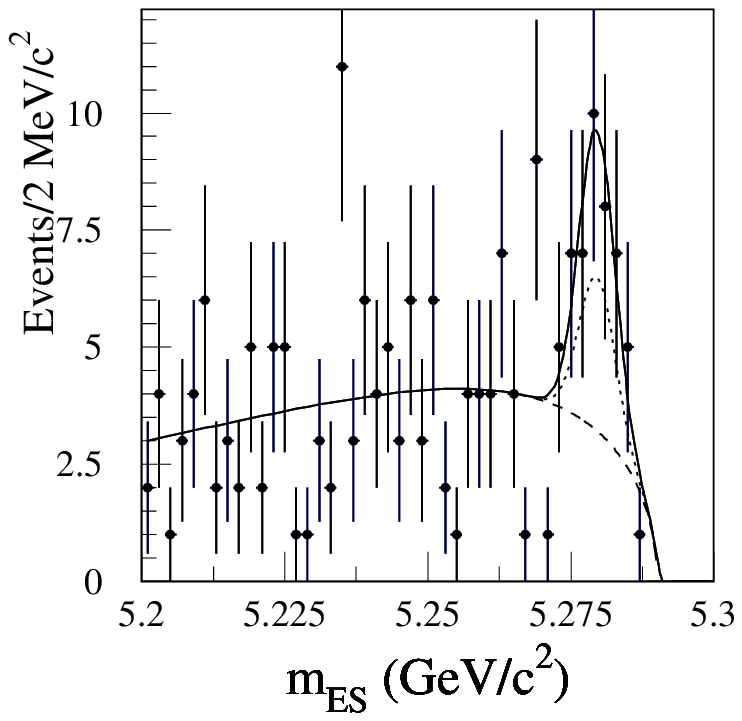}

  \includegraphics[width=0.15\textwidth]{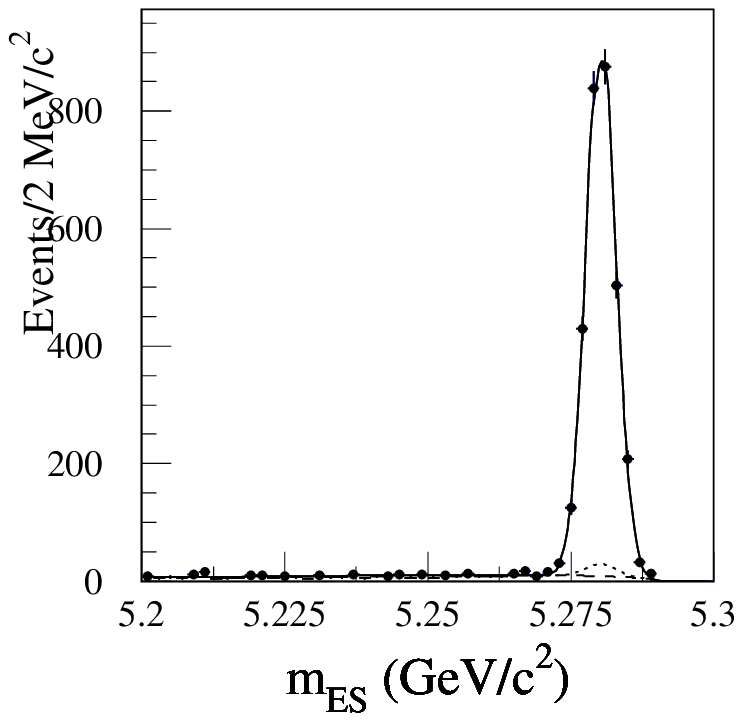}
  \includegraphics[width=0.15\textwidth]{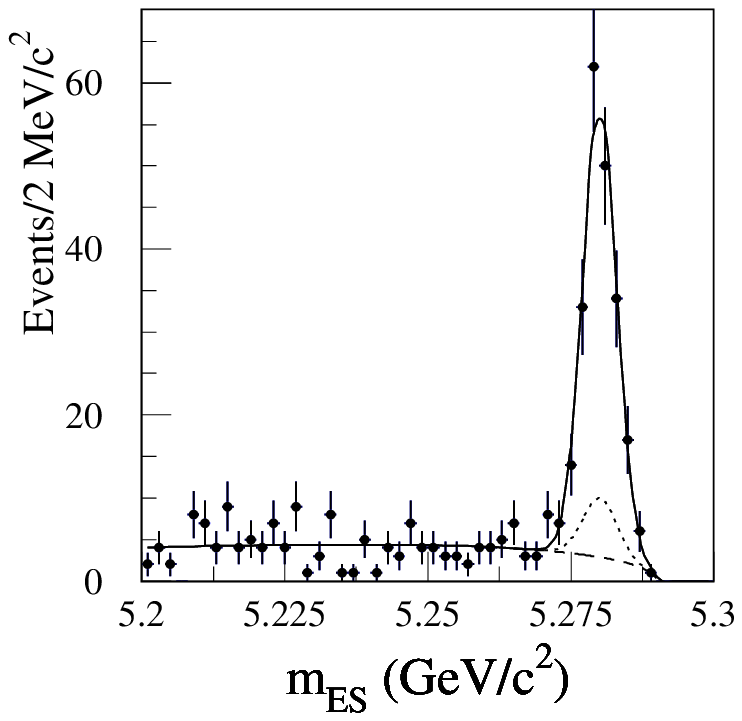}
  \includegraphics[width=0.15\textwidth]{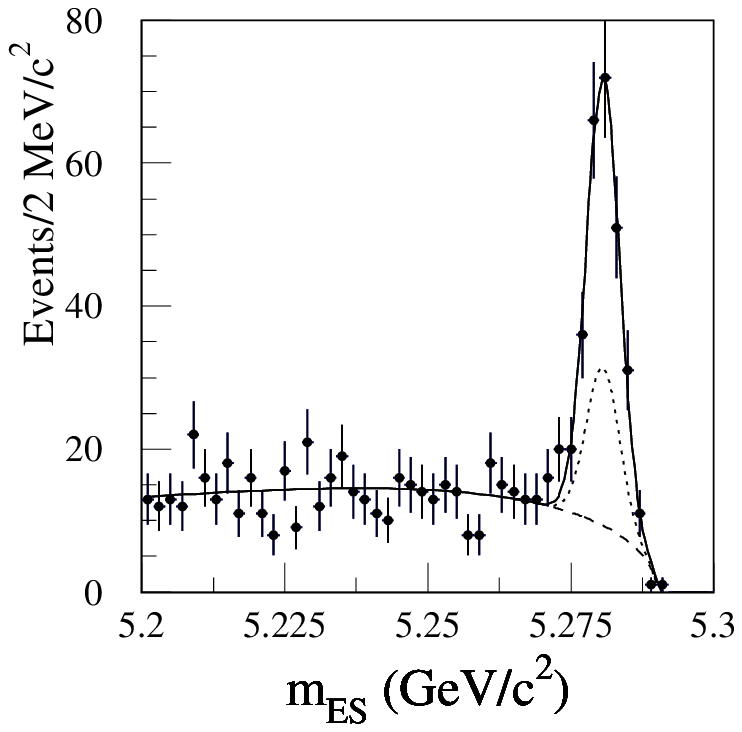}

  \includegraphics[width=0.15\textwidth]{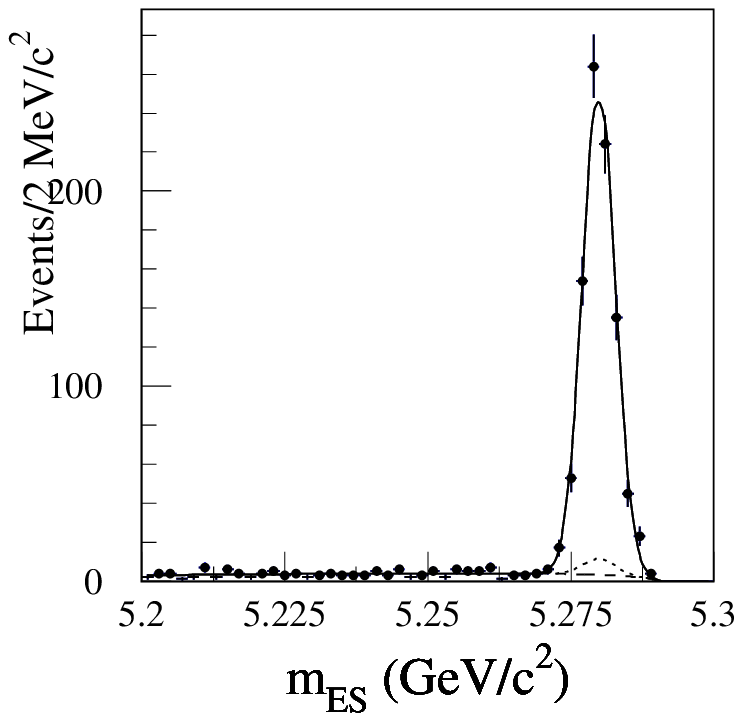}
  \includegraphics[width=0.15\textwidth]{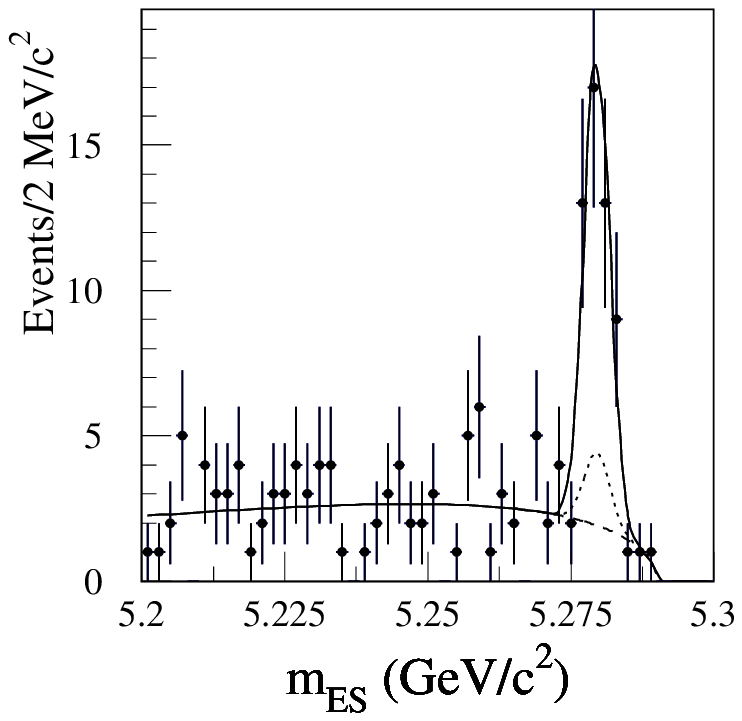}
  \includegraphics[width=0.15\textwidth]{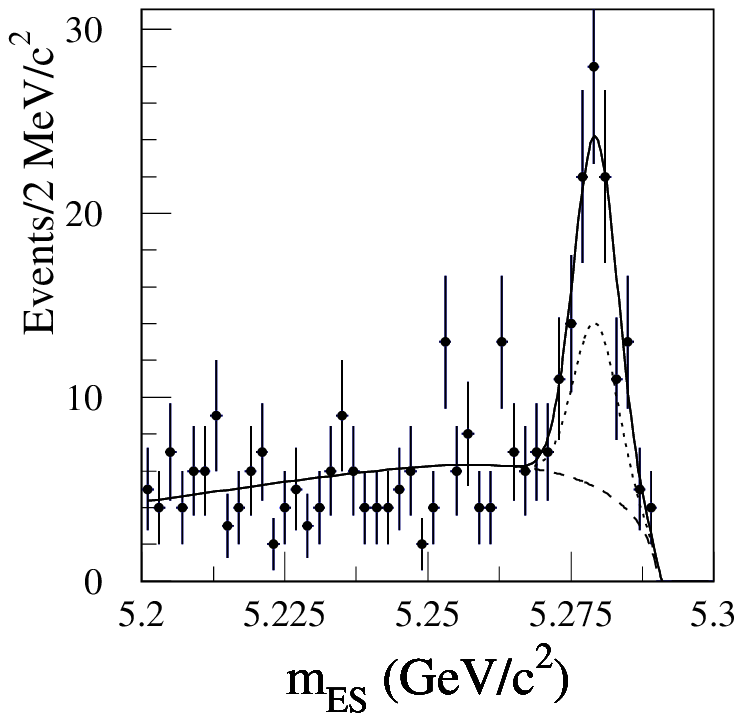}

  \includegraphics[width=0.15\textwidth]{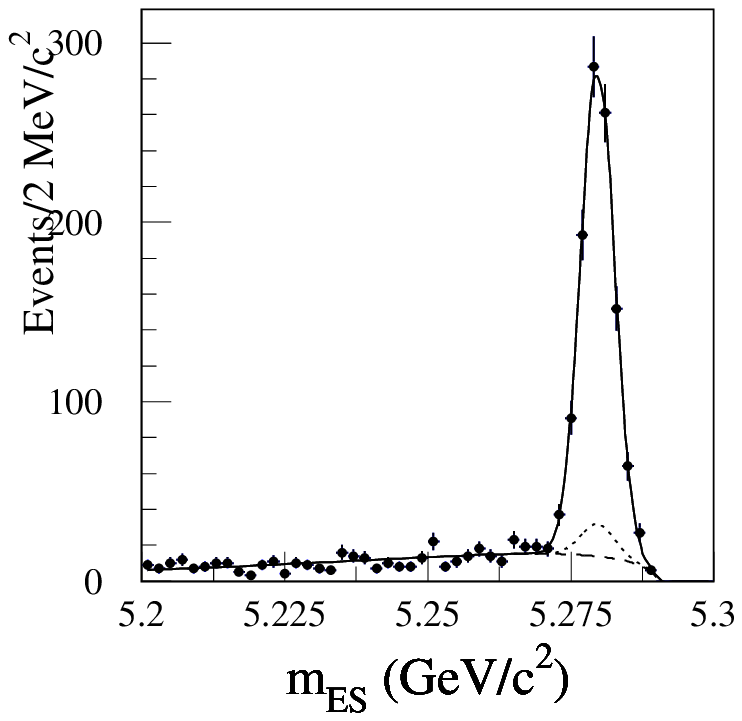}
  \includegraphics[width=0.15\textwidth]{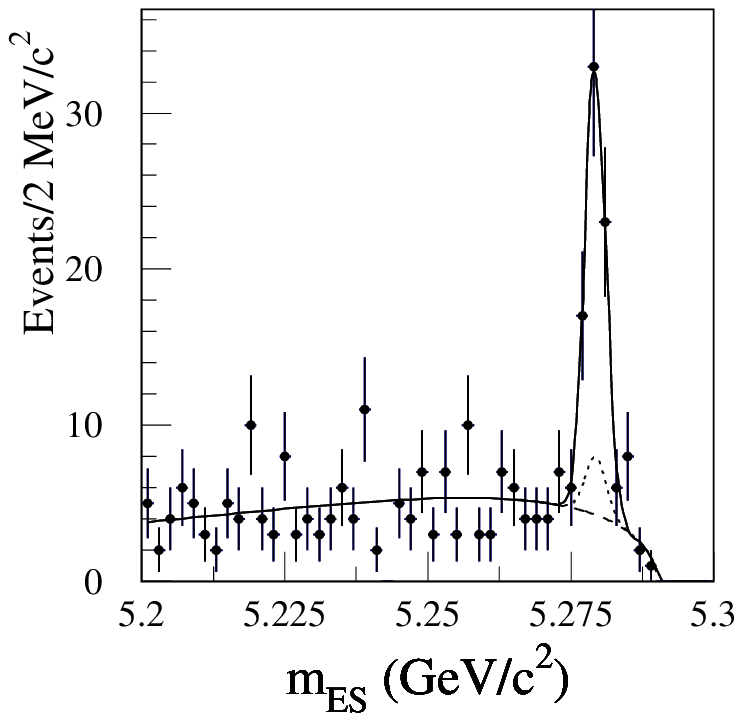}
  \includegraphics[width=0.15\textwidth]{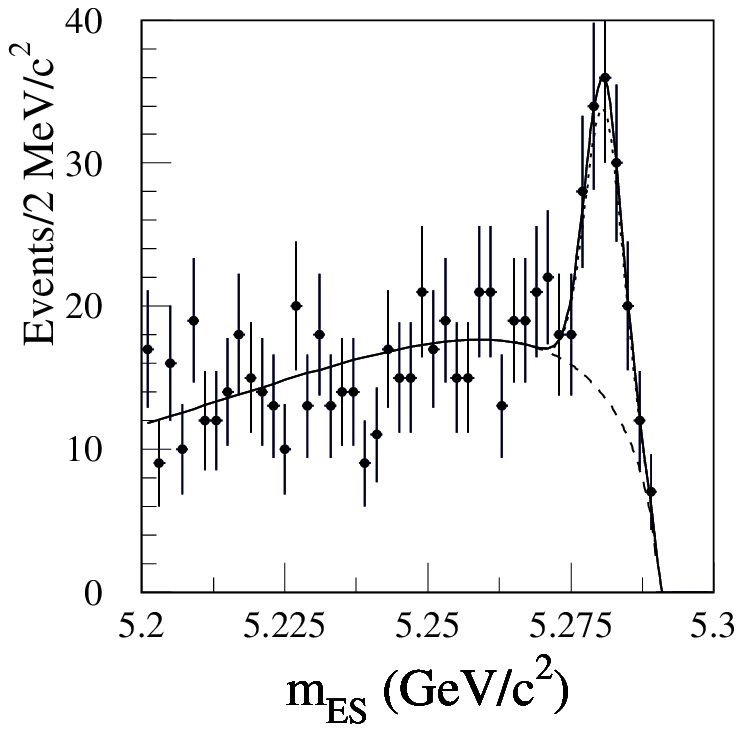}

  \includegraphics[width=0.15\textwidth]{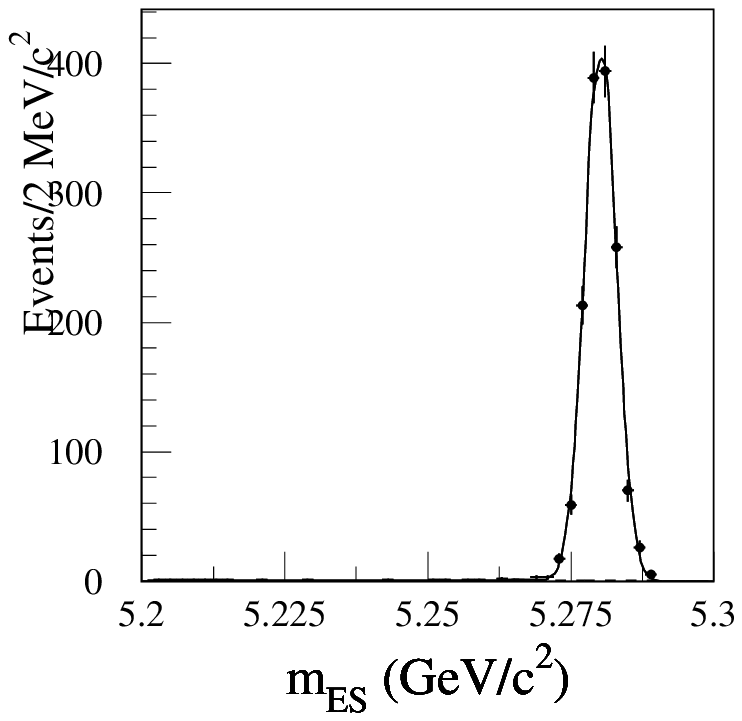}
  \includegraphics[width=0.15\textwidth]{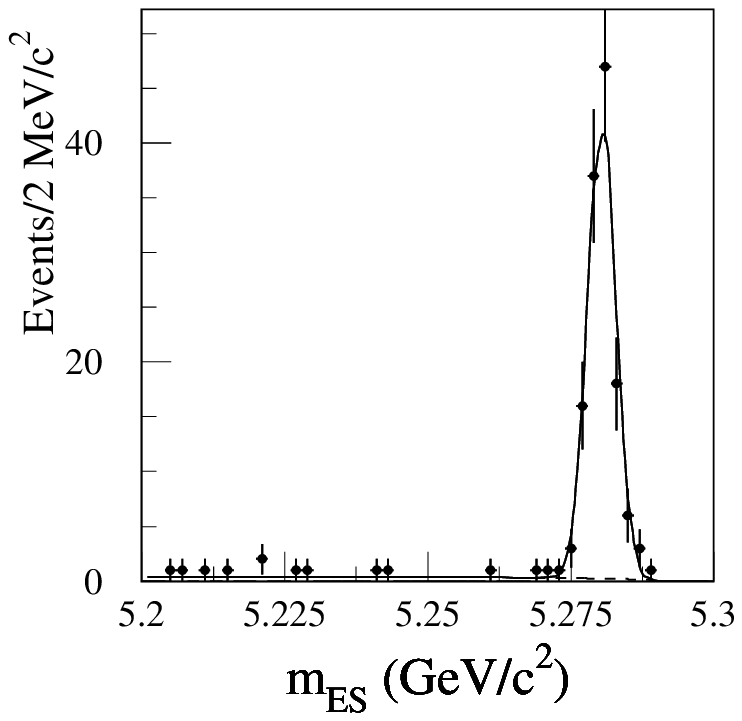}
  \includegraphics[width=0.15\textwidth]{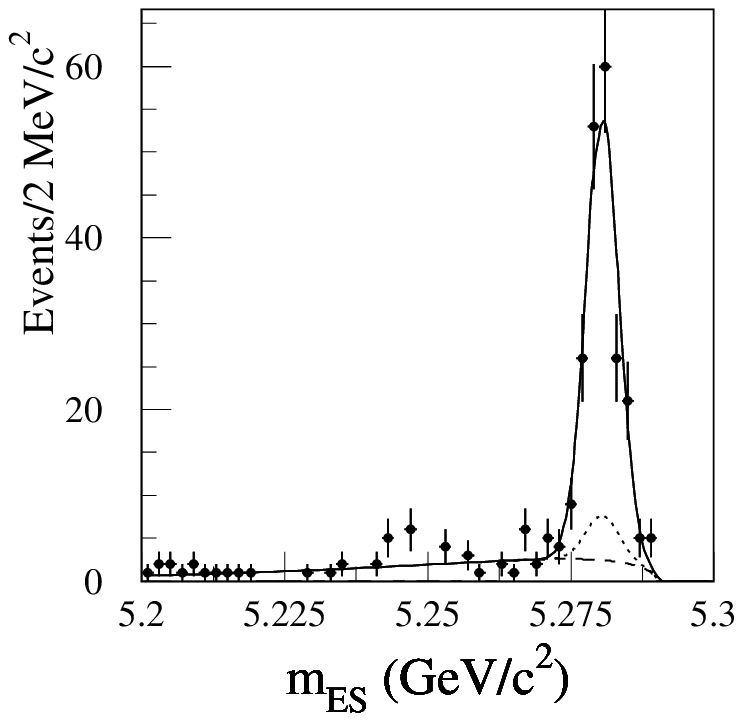}

  \includegraphics[width=0.15\textwidth]{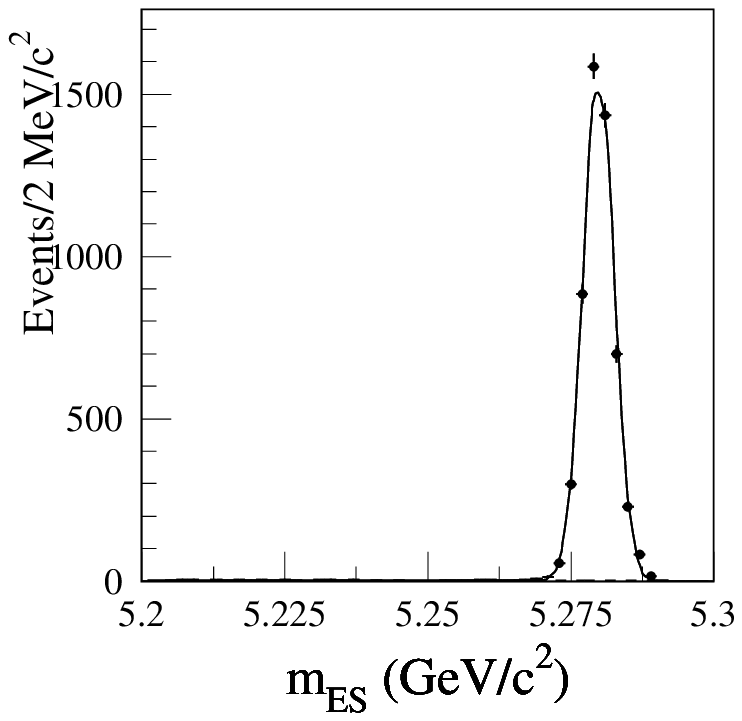}
  \includegraphics[width=0.15\textwidth]{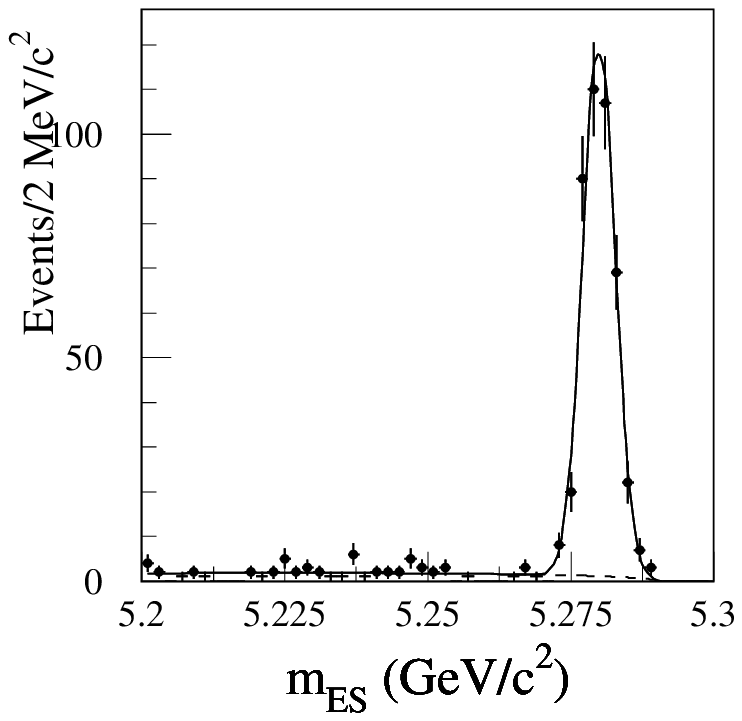}
  \includegraphics[width=0.15\textwidth]{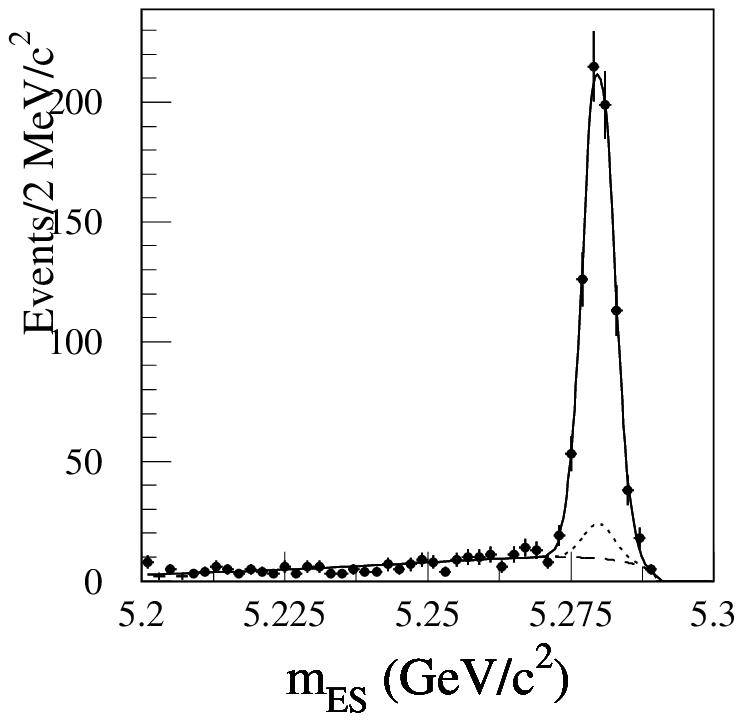}

 \caption{\label{fig:mes}
\mes distributions and fits within the \DeltaE\ signal region. 
From left to right, the columns show the distributions for the \jpsi, \psitwos and 
\chicone channels. From top to bottom, the rows show the distributions for the 
\B \ra ($c \bar{c}$) \Kstarz(\KS \piz), 
\B \ra ($c \bar{c}$) \Kstarz(\Kp \pim), 
\B \ra ($c \bar{c}$) \Kstarp(\KS \pip), 
\B \ra ($c \bar{c}$) \Kstarp(\Kp \piz), 
\B \ra ($c \bar{c}$) \KS and 
\B \ra ($c \bar{c}$) \Kp  decay modes. The dashed and dotted lines show the combinatorial and 
peaking backgrounds, respectively.}}
\end{figure}

The \B candidates are reconstructed by combining charmonium and kaon 
candidates and are characterized by two kinematic variables: the difference 
between the reconstructed energy of the \B candidate and the beam energy in the 
center-of-mass frame $\DeltaE = E_B^*-E_{beam}^*$, and the beam energy-substituted 
mass \mes, defined as $\mes \equiv \sqrt{E_{beam}^{*2}-{\bf p}_B^{*2}}$, 
where the $^*$ refers to quantities in the center-of-mass and ${\bf p}_B$ is the \B momentum. 
For a correctly reconstructed \B meson,
\DeltaE is expected to peak at zero and the energy-substituted mass \mes 
at the \B meson mass, 5.279 \gevcc. Only one reconstructed \B meson is 
allowed per event. For events that have multiple candidates, the candidate having 
the smallest $|\DeltaE|$  is chosen. The analysis is performed 
in a region of the \mes vs \DeltaE plane defined by $5.2 < \mes < 5.3$ \gevcc and 
$-0.12 < \DeltaE < 0.12$ \gev. 
A \DeltaE channel-dependent signal region 
is subsequently defined. The \mes distributions  within the \DeltaE\ signal 
region for candidate events are shown on Figure \ref{fig:mes}.

There are two components to the residual background in the \DeltaE signal region: the 
combinatorial background and a peaking component (component of the background that 
peaks at the same values of \DeltaE and \mes as the signal). The number of signal 
events $N_S$ is determined from the number of candidate events, $N_{\rm cand}$, after 
subtracting the peaking background. For this purpose, the \mes distribution within the 
\DeltaE signal region is fitted to the sum of an ARGUS function \cite{ref:argus}, which models 
the combinatorial background, and a Gaussian function. The value of $N_{\rm cand}$ is given 
by the integral of the Gaussian component. 
There are two contributions to the peaking background. The first is the cross-feed 
component that is due to 
\B \ra ($c \bar{c}$) \Kstar events from the three channels other than the one under 
consideration. 
The second contribution is from other \B decays with a \jpsi or 
\psitwos in the final state. 
To determine the extent of peaking background, the \mes distribution for simulated \BB 
events is fitted within the \DeltaE signal region by an ARGUS function and a Gaussian 
function. The  peaking background is taken as the integral of the Gaussian portion.

The branching fractions are obtained as:

\begin{equation}
  BF \equiv {N_{S} \over N_{\BB} \times \epsilon \times f} %\nonumber
 ,
\end{equation}

\noindent
where $N_{\BB}$ is the number of \BB events, $\epsilon$ is the selection 
efficiency and $f$ is the total secondary branching fraction. 
For channels with a \Kstar in the final state, the cross-feed contribution depends on 
the branching fractions that are being measured. It was estimated by an iterative procedure 
and found to be small. 

The systematic errors arise from the uncertainty on the number of \BB events (1.1$\%$), 
the secondary branching fractions (taken from Ref. \cite{ref:pdg2004}), the estimate of the 
selection efficiency, and the knowledge of the background. 
For the tracking efficiency, an error of 1.3$\%$ per track has been used. 
For the particle identification efficiency, the systematic error varies between 0.2$\%$ and 
3.7$\%$. 
The uncertainty on the detection and energy measurement of photons is 2.5$\%$, common 
to all channels, plus a small channel-dependent correction. The uncertainty 
on the \piz reconstruction is 5.0$\%$ for all channels, plus a 
channel-dependent correction.  
The overall selection efficiency depends on the angular distribution. 
The efficiency can be written as $\epsilon=a+A_0 b$, where $a$ and $b$ 
are obtained from the \Kstar helicity angle by 
$a = 3/4 \int (1-\cos^2\theta_{\Kstar}) \epsilon(\theta_{\Kstar}) 
\sin(\theta_{\Kstar})d\theta_{\Kstar}$ and 
$b = 3/4 \int (3\cos^2\theta_{\Kstar}-1) \epsilon(\theta_{\Kstar}) 
\sin(\theta_{\Kstar})d\theta_{\Kstar}$, 
and $A_0$ is the 
fraction of the longitudinal \Kstar polarization \cite{ref:a0}. The values of $a$ and $b$ are 
obtained from simulation. A systematic error 
varying from 3.4$\%$ to 8.6$\%$ has been derived for the branching fractions.  
In the default fit, the shape parameter of the ARGUS function is not constrained. 
To determine a systematic error due to the combinatorial background, 
a second fit with the shape parameter of the ARGUS function fixed to the 
value obtained from fitting the data in the \DeltaE sideband region was performed. 
The systematic uncertainty on the combinatorial background has been taken as 50$\%$ 
of the difference between the number of events obtained 
from the default fit and from the second fit. For the cross-feed 
component to the peaking background, the uncertainty of the corresponding branching 
fractions, taken from Ref. \cite{ref:pdg2004}, has been assigned as the systematic error. 
For the contribution coming from other \B decays with a \jpsi or a \psitwos in 
the final state, a 50$\%$ error has been assigned, accounting for the poor knowledge 
of the branching fractions of the contributing decay modes. 
Overall, the dominant contribution to the systematic error is from the secondary-decay 
branching fractions in the case of \psitwos $K^{(*)}$ and \chicone $K$, the efficiency
determination in the case of \jpsi $K^{(*)}$, and the background subtraction in the
case of \chicone \Kstar.

\begin{table}[!htb]
\caption{\label{tab:results}
Measured branching fractions for exclusive decays of \B mesons to 
charmonium and kaon final states. The first error is statistical and the second 
systematic.}
\begin{center}
\begin{tabular}{lc}  \hline \hline
Channel & Branching fraction ($\times 10^{-4}$) \\ \hline

\Bz \ra \jpsi \Kstarz      &  13.09$\pm$0.26$\pm$0.77 \\ 
\Bp \ra \jpsi \Kstarp      &  14.54$\pm$0.47$\pm$0.97 \\ 
\Bp \ra \jpsi \Kp          &  10.61$\pm$0.15$\pm$0.48 \\ 
\Bz \ra \jpsi \Kz          &  8.69$\pm$0.22$\pm$0.30 \\ 

\Bz \ra \psitwos \Kstarz   &  6.49$\pm$0.59$\pm$0.97 \\ 
\Bp \ra \psitwos \Kstarp   &  5.92$\pm$0.85$\pm$0.89 \\ 
\Bp \ra \psitwos \Kp       &  6.17$\pm$0.32$\pm$0.44 \\ 
\Bz \ra \psitwos \Kz       &  6.46$\pm$0.65$\pm$0.51 \\ 

\Bz \ra \chicone \Kstarz   &  3.27$\pm$0.42$\pm$0.64 \\ 
\Bp \ra \chicone \Kstarp   &  2.94$\pm$0.95$\pm$0.98  \\
\Bp \ra \chicone \Kp       &  5.79$\pm$0.26$\pm$0.65 \\ 
\Bz \ra \chicone \Kz       &  4.53$\pm$0.41$\pm$0.51 \\ \hline

\end{tabular}
\end{center}
\end{table}

The branching fraction measurements are summarized in Table \ref{tab:results}. 
We have taken $R^{+/0}$, the ratio 
$\Gamma$($\Upsilon(4S) \ra \BpBm$)/$\Gamma$($\Upsilon(4S) \ra \BzBzb$), to be 
unity for the determination of the branching fractions. 
Assuming isospin invariance in the  \B \ra charmonium \kaon (\Kstar) 
decays, we can compute $R^{+/0}$. Using the ratio of the charged 
to neutral \B meson lifetimes 
$\tau_{\Bp} / \tau_{\Bz} = 1.086 \pm 0.017$ \cite{ref:pdg2004}, we obtain:

\begin{equation}
R^{+/0} \equiv { \Gamma( \Upsilon(4S) \ra \BpBm)  \over \Gamma( \Upsilon(4S) \ra \BzBzb ) } = 
1.06\pm0.02\pm0.03, 
%\nonumber
\end{equation}

\noindent
where the first error is statistical and the second systematic. 
The ratios of the branching fractions for $\B \ra (c \bar{c}) \Kstar$ and 
$\B \ra (c \bar{c}) K$ for the three $(c \bar{c})$ states are presented in 
Table \ref{tab:vectscal}. 
For each of the charmonium states, the average of the charged and neutral 
measurements is also shown.

\begin{table}[!htb]
\caption{\label{tab:vectscal}
Results for ratios of the branching fractions for charmonium and a \Kstar versus 
charmonium and a \kaon. The first error is statistical and the second systematic. }
\begin{center}
\begin{tabular}{lr} \hline \hline
Ratio  &  Result \\ \hline
$\BR(\Bz \ra \jpsi \Kstarz)/\BR(\Bz \ra \jpsi \Kz)$       &  1.51$\pm$0.05$\pm$0.08  \\ 
$\BR(\Bp \ra \jpsi \Kstarp)/\BR(\Bp \ra \jpsi \Kp)$       &  1.37$\pm$0.05$\pm$0.08  \\ 
$\BR(\B \ra \jpsi \Kstar)/\BR(\B \ra \jpsi \kaon)$        &  1.44$\pm$0.04$\pm$0.06  \\ \hline 

$\BR(\Bz \ra \psitwos \Kstarz)/\BR(\Bz \ra \psitwos \Kz)$ &  1.00$\pm$0.14$\pm$0.09  \\ 
$\BR(\Bp \ra \psitwos \Kstarp)/\BR(\Bp \ra \psitwos \Kp)$ &  0.96$\pm$0.15$\pm$0.09  \\ 
$\BR(\B \ra \psitwos \Kstar)/\BR(\B \ra \psitwos \kaon)$  &  0.98$\pm$0.10$\pm$0.07  \\ \hline 

$\BR(\Bz \ra \chicone \Kstarz)/\BR(\Bz \ra \chicone \Kz)$ &  0.72$\pm$0.11$\pm$0.12  \\ 
$\BR(\Bp \ra \chicone \Kstarp)/\BR(\Bp \ra \chicone \Kp)$ &  0.51$\pm$0.17$\pm$0.16  \\ 
$\BR(\B \ra \chicone \Kstar)/\BR(\B \ra \chicone \kaon)$  &  0.65$\pm$0.09$\pm$0.10  \\ \hline
\end{tabular}
\end{center}
\end{table}

Finally, we have measured the charge asymmetries 

\begin{equation}
A \equiv { \BR(\Bp \ra (c \bar{c}) K^{+(*)}) -  \BR(\Bm \ra  (c \bar{c}) K^{-(*)}) \over 
  \BR(\Bp \ra (c \bar{c}) K^{+(*)}) +  \BR(\Bm \ra  (c \bar{c}) K^{-(*)}) }, %\nonumber
\end{equation}

\noindent
using efficiencies determined separately for the two charges. The results are presented 
in Table \ref{tab:ca}. No statistically significant asymmetry is observed.

\begin{table}[!t]
\caption{\label{tab:ca}
Results for charge asymmetries. The first error is statistical and the second systematic. }
\begin{center}
\begin{tabular}{lr} \hline \hline
Final state       &  Asymmetry \\ \hline
 \jpsi \Kp        &  -0.030$\pm$0.014$\pm$0.010  \\ 
 \jpsi \Kstarp    &   0.048$\pm$0.029$\pm$0.016  \\ 
 \psitwos \Kp     &   0.052$\pm$0.059$\pm$0.020  \\ 
 \psitwos \Kstarp &  -0.077$\pm$0.207$\pm$0.051  \\ 
 \chicone \Kp     &   0.003$\pm$0.076$\pm$0.017  \\ 
 \chicone \Kstarp &  -0.471$\pm$0.378$\pm$0.268  \\ \hline
\end{tabular}
\end{center}
\end{table}

In summary, branching fraction measurements of exclusive 
\B decays to charmonium (\jpsi, \psitwos and \chicone) and \kaon or \Kstar have been presented. 
Our results for \jpsi and \psitwos are in good agreement 
with previous measurements \cite{ref:pdg2004} and exhibit comparable or superior precision. 
Our \chicone results have much better precision than earlier measurements. The 
\Bp \ra \chicone \Kstarp mode was previously unmeasured. 
Assuming isospin invariance, we find 
the ratio of charged- to neutral-\B meson production on the $\Upsilon$(4S) 
resonance to be compatible with unity within 1.7 standard deviations. No 
direct \CP violation has been observed in the charge asymmetries.

We are grateful for the excellent luminosity and machine conditions provided by our \pep2\ 
colleagues, and for the substantial dedicated effort from the computing organizations 
that support \babar. The collaborating institutions wish to thank SLAC for its support 
and kind hospitality. This work is supported by DOE and NSF (USA), NSERC (Canada), IHEP (China), 
CEA and CNRS-IN2P3 (France), BMBF and DFG (Germany), INFN (Italy), FOM (The Netherlands), 
NFR (Norway), MIST (Russia), and PPARC (United Kingdom). Individuals have received support 
from CONACyT (Mexico), A.~P.~Sloan Foundation, Research Corporation, and 
Alexander von Humboldt Foundation.


\begin{thebibliography}{99}

\bibitem{ref:oldanalysis}
\babar\ Collaboration, B. Aubert {\em et al.}, 
Phys. Rev. D. {\bf 65}, 032001 (2002).


\bibitem{ref:smallasymm}
T. Brown, S. Pakvasa and S.F. Tuan, Phys. Lett. {\bf B136}, 117 (1984); 
I. Dunietz, Phys. Lett. {\bf B316}, 561 (1993).


\bibitem{ref:belleca}
BELLE Collaboration, K. Abe {\em et al.}, 
Phys. Rev. D. {\bf 67}, 032003 (2003).

\bibitem{ref:babar}
\babar\ Collaboration, B. Aubert {\em et al.},
Nucl. Instr. and Methods {\bf A 479}, 1 (2002).

\bibitem{ref:lat}
A. Drescher {\em et al.}, Nucl. Instrum. Methods {\bf A237}, 464 (1985).

\bibitem{ref:zernike}
R. Sinkus and T. Voss, Nucl. Instrum. Methods {\bf A391}, 360 (1997).

\bibitem{ref:argus}
ARGUS Collaboration, H. Albrecht {\em et al.}, 
Z. Phys. C {\bf 48}, 543 (1990).

\bibitem{ref:pdg2004}
Particle Data Group, 
S. Eidelman {\em et al.}, Phys. Lett. {\bf B592}, 1 (2004).

\bibitem{ref:a0}
\babar\ Collaboration, B. Aubert {\em et al.},
Phys. Rev. Lett. {\bf 87}, 241801 (2001).


\end{thebibliography}
\end{document}